\documentclass[preprint,aps]{revtex4}
\usepackage{euscript}
\usepackage{amsmath}
\usepackage{epsfig}
\begin{document}
\preprint{UCI-TR-2009-15}
\title{Model with Possible Fields Generated by Higher Dimensional Super Conducting Cosmic Strings}
\author{Aaron Roy\footnote{Electronic address:roya@uci.edu}
%\addtocounter{footnote}{3}%
}
\affiliation{
Department of Physics and Astronomy, University of California, Irvine,
California 92697-4575}%\PRE{\vspace*{.5in}} }

\begin{abstract}
In this paper we investigate the theory behind the results in [11]. In [11] we calculated the magnetic dipole moment of the muon and the electric dipole moments of the muon, electron and the neutron (in a simple quark model) to first order in loop corrections in both S$_1$ and S$_1 \backslash \, \boldsymbol{Z}_2$. In these calculations in [11] we investigated the effect of fields possibly generated by higher dimensional superconducting cosmic strings [12] that interact with the charged fields on the manifold. In comparing the results in [11] with standard model precision tests for the electric and magnetic dipole moments of the various fermions in the model, we were able to obtain upper limits on the compactification size as well as an upper limit for the new $b$ parameter. This new model has several important phenomenological implications. One of these is a theoretical phenomenon that is a source for parity violation in QED processes. In this paper we will be presenting the theory of the model in the compactification $M_4 \, \bigotimes$ S$_1 \backslash \, \boldsymbol{Z}_2$ (orbifold geometry). The theory predicts nontrivial couplings of the Higgs and lepton fields to the SU(2) gauge bosons, these differ from the standard model couplings. The theory also expounds upon the standard model results for the masses of the charged fields in the model and has other significant physical implications. The model is rooted in the notion that very light charged particles traveling next to superconducting cosmic strings [12] at distances on the order of the compactification size of the extra dimensional space, could generate currents that intern can create magnetic fields that interact with the particle fields on the UED manifold. Please see section V of [11] for a more detailed discussion of this.     
\end{abstract}

\maketitle

\section{Introduction}
In this paper we examine the consequences of magnetic fields that could be produced by light charged particles traveling in proximity to superconducting cosmic strings [12], where the separation scale between the particle and the string is on the order of the compactification size of S$_1 \backslash \, \boldsymbol{Z}_2$ [1,2,3]. These external magnetic fields will have some flux associated with the manifold of the model $M_4 \, \bigotimes$ S$_1 \backslash \, \boldsymbol{Z}_2$. This means that the magnetic fields responsible for the fluxes would, at the very least, propagate in 6-D space while our fields of the model are confined to the 5-D manifold. We are not concerned with the geometry of the higher dimensional space that these fluxes exist in, only their affects on the fields in our model in $M_4 \, \bigotimes$ S$_1 \backslash \, \boldsymbol{Z}_2$. 

These fluxes affect the charged fields with nontrivial periodicities, where in general, these nontrivial periodicities are not simply a shift in mode number. With these fluxes, a new parameter is introduced, the flux parameter. This addition to an SU(2)$\bigotimes$U(1) electroweak model in $M_4 \, \bigotimes$ S$_1 \backslash \, \boldsymbol{Z}_2$ provides a novel mechanism for parity violation in QED processes and thus affects the EDM's of various charged fermions in the model [11]. The fluxes also allow for a new way of SU(2)$\bigotimes$U(1) symmetry breaking. In addition, we will find nontrivial couplings of the Higgs and lepton fields to the SU(2) gauge bosons due to the fluxes. The model we present here gives a physical source for these nontrivial periodicities. These nontrivial periodicities only affect the charged particles in the theory in a non arbitrary way, we do not add these nontrivial periodicities in an ad-hoc fashion. The fluxes introduced in this paper have several phenomenological implications with respect to experiment.  

We will present the masses of all the fields in the model in both the full 5-D space and also in the effective brane. The masses of the charged fields in the model will have flux dependence and in the zero mode limit these expressions expound upon the standard model results. Yet another interesting consequence of this new model is the solution to the degrees of freedom problem which arises when the charged $W$'s acquire a mass before any Higgs mechanism due to the external flux.

Another important implication of the theory is the Higgs mechanism itself. We will see that a more complicated gauge is required for the mechanism and that this gauge goes smoothly into the unitary gauge in the limit as the flux goes to zero. The exact closed form expression for the gauge does not exist. Addition of leptons and quarks to the model proves challenging due to the nontrivial periodicity of the fields as we will see in section IV. With this particular dimensionality we will also have unwanted 5th components of the gauge fields. These unwanted components are easily dealt with in the orbifold geometry. Finally we will present an estimate of the upper limit on the ratio of the flux and the compact radius $R$ of the model. 

In Sec. II we will introduce the model and incorporate the fluxes in the theory. In Sec. III we will discuss the Higgs mechanism for the model and present the masses for each of the fields along with the mass ratio relation. In Sec. IV we will add first generation leptons and quarks to the model and give their masses. We will conclude with an estimate for the upper limit of the ratio of the flux and compact radius for our particular manifold in Sec. V.       
\section{Five dimensional electroweak model in $M_4 \, \bigotimes$ S$_1 \backslash \, \boldsymbol{Z}_2$ \\ geometry}
\subsection{Theory of the model}
In the orbifold geometry $M_4 \, \bigotimes$ S$_1 \backslash \, \boldsymbol{Z}_2$, we may identify the points $0$ and $\pi$ on the compact circle S$_1$ as topologically distinct. This distinctness gives us an additional mathematical degree of freedom which can be taken advantage of in the Fourier expansions of the fields in the model. For an arbitrary field $\psi(x^\mu ,y)$ where $y$ is the extra coordinate (which will be defined here as an arc length along the circle 
S$_1$ in S$_1 \backslash \, \boldsymbol{Z}_2$) we have
\begin{equation}
\psi(x^\mu ,y) = \frac{1}{\sqrt{2\pi R}}\sum _{n \, = \, -\infty} ^\infty \psi _n(x^\mu) e^{iny/R} \, , \nonumber 
\end{equation} 
which under the inversion \, $y \rightarrow y + \pi R$ \, gives 
\begin{equation}
\psi(x^\mu ,y+\pi R) = \frac{1}{\sqrt{2\pi R}}\sum _{n \, = \, -\infty} ^\infty (-1)^n \psi _n(x^\mu) e^{iny/R} \, . \nonumber 
\end{equation}

With the extra mathematical degree of freedom afforded by the topologically distinct points $0$ and $\pi$ on the orbifold, we can now assign even and odd nature to the fields under \,  $y \rightarrow y + \pi R$ \,. Therefore, we choose the fifth components of the gauge fields in the model to have odd parity under \, $y \rightarrow y + \pi R$ \, (and hence only have odd mode numbers) and all of the other fields to be even (will only contain even mode numbers). What this does is decouple the 5th components of the gauge fields from the rest of the fields in the model. This can easily be seen once the above Fourier expansions are used in the Lagrangian density and the extra coordinate is integrated over in the action. Then any interaction term involving the 5th components of the gauge field modes and any other field modes will vanish due to the orthogonality relation \, $\frac{1}{2\pi R}\int _0 ^{2\pi R}e^{i(n-m)y/R}\,dy = \delta _{n,m}$ \, since any sum of even mode numbers with an odd mode number can never satisfy the above orthogonality relation, giving zero for the RHS. Of course these phenomenologically forbidden 5th components still carry degrees of freedom in the model, they just don't interact with the other fields.    

We have the following 5-D lagrangian density for our SU(2)$\bigotimes$U(1) electroweak model in $M_4 \, \bigotimes$ S$_1 \backslash \, \boldsymbol{Z}_2$
\begin{equation}
\EuScript{L} = (D_A \varphi )^\dagger (D^A \varphi) - \frac{1}{2}Tr(F_{AB}F^{AB}) - \frac{1}{4}f_{AB}f^{AB} + \mu ^2 \varphi ^\dagger \varphi - \frac{\lambda}{2}(\varphi ^\dagger \varphi )^2
\end{equation}
where $A,B=0,1,2,3,5$ with each field being a function of $x^\mu ,y$ where again, $y$ is the extra coordinate in our system and will be defined as an arc length as stated earlier. 
We will assume a flat metric, 
\begin{equation}
g^{AB}=
  \begin{cases} 
    0& \text{if $A\neq B$},\\
   -1& \text{if $A=B=1,2,3,5$},\\ 
    1& \text{if $A=B=0$}. 
  \end{cases} 
\end{equation} 
As usual $D_A = \partial _A + igW_A + \frac{i}{2}g'B_A$ with 
\begin{equation}
W_A = W_A ^i \frac{\tau ^i}{2} = \begin{pmatrix} \frac{1}{2}W_A ^3 & \frac{1}{\sqrt{2}}W_A ^+ \\ \frac{1}{\sqrt{2}}W_A ^- & - \frac{1}{2}W_A ^3 \end{pmatrix}\, ,  
\end{equation}
\begin{equation}
F_{AB} = \partial _A W_B - \partial _B W_A - ig[W_A , W_B].
\end{equation}  
\subsection{Inclusion of external magnetic flux}
As discussed in the introduction, for very light charged particles traveling next to superconducting cosmic strings [12] at distances on the order of the compactification size of the extra dimensional space, currents can be generated that intern can create magnetic fields that interact with the particle fields on the UED manifold. Please see section V and also section V of [11] for a calculation of this phenomenon. These fields will have a magnetic flux associated with the  manifold, this results in a magnetic flux threading through the extra dimension (perpendicular to the extra coordinate, which would be the case for very small $R$). These fluxes interact with fields on the manifold in which all of the charged fields will be affected by the contribution $e^{iQby/R}$ where $Q$ is the charge of the field affected by the flux (i.e. $Q = -1$ for the electron etc.) and $b = \frac{e}{\hbar c}\times \rm {flux}$ (Gaussian units). Please note that in this paper we will be using Heaveside-Lorentz units exclusively ($\hbar = c = 1$). Under the orbifold inversion \, $y \rightarrow y + \pi R$ \, we now have for an arbitrary field $\psi(x^\mu ,y)$ with charge $Q$,
\begin{equation}
\psi(x^\mu ,y+\pi R) = e^{i\pi Q b}\frac{1}{\sqrt{2\pi R}}\sum _{n \, = \, -\infty} ^\infty (-1)^n \psi _n(x^\mu) e^{i(n+Qb)y/R} \, . \nonumber 
\end{equation}

The nontrivial periodicity as a result of the flux does not affect the even and odd nature of the field modes as described in the previous section. The reason is because the flux parameter $b$ is a approximately a constant for very small compactification size ($R$) which is required for consistency of precision tests of the standard model at the current energies being probed [13]. The flux would be proportional to the flat surface area mapped out by the extra coordinate in our geometry and hence would be a constant for very small $R$. Therefore the fields chosen with even modes would have the same nontrivial periodicity factor (constant flux) out in front of their Fourier expansions under \, $y \rightarrow y + \pi R$ \, as do the odd modes choice, thus preserving the even and odd nature of the fields afforded by the orbifold geometry in the model.       

The charge assignments for the model are standard, if we let $\phi = \begin{pmatrix} \phi ^1 \\ \phi ^2\end{pmatrix}$ then $\phi ^1$ will have charge $Q = +1$ and $\phi ^2$ will be neutral.
Therefore with the flux only the top component of $\phi$ is affected as follows: \\
$\phi \underset{\rm{flux}}{\rightarrow} \begin{pmatrix} e^{iby/R} & 0 \\ 0 & 1\end{pmatrix}\phi$ \, and if we define \, $B(y) = \begin{pmatrix} e^{iby/R} & 0 \\ 0 & 1\end{pmatrix}$ \, then  
\begin{equation}
\phi \underset{\rm{flux}}{\rightarrow}B\phi 
\end{equation} 
and for the $W_A$, 
\begin{equation} 
W_A \underset{\rm{flux}}{\rightarrow} \\ \begin{pmatrix} \frac{1}{2}W_A ^3 & \frac{1}{\sqrt{2}}e^{iby/R}W_A ^+ \\ \frac{1}{\sqrt{2}}e^{-iby/R}W_A ^- & - \frac{1}{2}W_A ^3\end{pmatrix} \nonumber
\end{equation}
or  
\begin{equation}
W_A \underset{\rm{flux}}{\rightarrow} B W_A B^\dagger .                                     
\end{equation}
Notice that we now have nontrivial couplings of the Higgs field to the SU(2) gauge bosons as a result of the fluxes. We also have  
\begin{equation}
{\rm{Tr}}(F_{\mu \nu}F^{\mu \nu})\underset{\rm{flux}}{\rightarrow}{\rm{Tr}}(F_{\mu \nu}F^{\mu \nu}) \, ,  \nonumber
\end{equation}
along with
\begin{equation}
F_{\mu 5} \underset{\rm{flux}}{\rightarrow} BF_{\mu 5}B^\dagger - (\partial _y B) W_\mu B^\dagger - B W_\mu (\partial _y B^\dagger) \, , 
\end{equation} 
which gives
\begin{equation}
{\rm{Tr}}(F_{\mu 5}F^{\mu 5})\underset{\rm{flux}}{\rightarrow}\frac{1}{2}F^3 _{\mu 5}F^{3\mu 5} - (\partial _\mu W_5 ^- - \partial _y W_\mu ^- + \frac{ib}{R}W_\mu ^- )(\partial ^\mu W_5 ^+ - \partial _y W^{\mu +} - \frac{ib}{R}W^{\mu +}) \nonumber
\end{equation}
\begin{equation}
+ \rm{(cubic \, and \, quartic \, terms)} \nonumber.
\end{equation}

We can see that with the flux, the charged $W_\mu$'s pick up a mass $m_w = \frac{|b|}{R}$. Let us redefine the charged $W_\mu$ fields as follows: 
\begin{eqnarray}
\tilde{W}^- _\mu = W^- _\mu - \Lambda \partial _\mu W_5 ^- \\
\tilde{W}^+ _\mu = W^+ _\mu - \beta \partial _\mu W_5 ^+
\end{eqnarray}  
which finally gives
\begin{equation}
{\rm{Tr}}(F_{\mu 5}F^{\mu 5})\underset{\rm{flux}}{\rightarrow}\frac{1}{2}F^3 _{\mu 5}F^{3\mu 5}  - ( \partial _y \tilde{W}_\mu ^- - \frac{ib}{R}\tilde{W}_\mu ^- )(\partial _y \tilde{W}^{\mu +} + \frac{ib}{R}\tilde{W}^{\mu +}) \nonumber
\end{equation}
\begin{equation}
+ \rm{(cubic \, and \, quartic \, terms)}
\end{equation} 
where $\Lambda = (\partial _y - \frac{ib}{R})^{-1}$ and $\beta = (\partial _y + \frac{ib}{R})^{-1}$. The reason that we made the field transformations in equation (10) will become clear in section IIIB. With the fluxes, it is obvious that the 
SU(2)$\otimes$U(1) symmetry is broken. 
\section{Higgs mechanism}
Lets now look at $(D_A \phi)^\dagger (D^A \phi)$ which gives 
\begin{equation}
(D_A \phi)^\dagger (D^A \phi)\underset{\rm{flux}}{\rightarrow}   
\phi ^\dagger B^\dagger[\overleftarrow{\partial} _A + ig B W_A B^\dagger + \frac{i}{2}g'B_A][\overrightarrow{\partial} ^A - ig B W^A B^\dagger - \frac{i}{2}g'B^A]B\phi \nonumber
\end{equation}
or
\begin{equation} 
 (D_A \phi)^\dagger (D^A \phi)\underset{\rm{flux}}{\rightarrow}   
\phi ^\dagger[\overleftarrow{\partial} _\mu + ig\tilde{W}_\mu + ig\partial _\mu T + \frac{i}{2}g'B_\mu][\overrightarrow{\partial} ^\mu - ig\tilde{W}^\mu - ig\partial ^\mu T - \frac{i}{2}g'B^\mu]\phi \nonumber
\end{equation}
\begin{equation}  
- \phi ^\dagger[\overleftarrow{\partial _y} + (\partial _y B^\dagger )B + igW_5 + \frac{i}{2}g'B_5][\overrightarrow{\partial _y} + B^\dagger (\partial _y B) - igW_5 - \frac{i}{2}g'B_5]\phi
\end{equation}
where 
\begin{equation}
W_\mu = \tilde{W}_\mu + \partial _\mu T \nonumber \, ,
\end{equation}
\begin{equation}
\tilde{W}_\mu = \begin{pmatrix} \frac{1}{2}W^3 _\mu & \frac{1}{\sqrt{2}}\tilde{W} ^+ _\mu \\ \frac{1}{\sqrt{2}}\tilde{W} ^- _\mu & -\frac{1}{2}W^3 _\mu\end{pmatrix} \, , \nonumber
\end{equation}
and
\begin{equation} 
T = \frac{1}{\sqrt{2}}\begin{pmatrix} 0 & \beta W^+ _5 \\ \Lambda W^- _5 & 0\end{pmatrix} \, , \nonumber 
\end{equation}
from the field redefinitions in (8) and (9). 

From equation (11) we have the term \, $-\phi ^\dagger (\partial _y B^\dagger)(\partial _y B)\phi = -\phi ^\dagger
\begin{pmatrix} \frac{b^2}{R^2} & 0 \\ 0 & 0\end{pmatrix}\phi$. If this is combined with the terms \,
$\mu ^2 \phi ^\dagger \phi - \frac{\lambda}{2}(\phi ^\dagger \phi )^2$ \, then define
\begin{equation}
V(\phi ^\dagger \phi) = -\phi ^\dagger \begin{pmatrix} \mu ^2 - \frac{b^2}{R^2} & 0 \\ 0 & \mu ^2\end{pmatrix}\phi + 
\frac{\lambda}{2}(\phi ^\dagger \phi )^2.
\end{equation} 
If we minimize the potential in equation (12) then there is one local minimum at 
$\langle \phi \rangle _{\rm{local \, minimum}} = \begin{pmatrix} \sqrt{\frac{\mu ^2 - \frac{b^2}{R^2}}{\lambda}} \\ 0\end{pmatrix}$
and a global minimum at $\langle \phi \rangle _0 = \begin{pmatrix} 0 \\ \sqrt{\frac{\mu ^2}{\lambda}} \end{pmatrix}$ and thus the "physical" vacuum state is $\langle \phi \rangle _0 = \begin{pmatrix} 0 \\ v\end{pmatrix}$ where \, 
$v = \sqrt{\frac{\mu ^2}{\lambda}}$ [4].

Parametrize the $\phi$ field as follows:
\begin{equation} 
\phi (x^\mu ,y) = \frac{1}{\sqrt{2}}U(x^\mu ,y)B(y)\begin{pmatrix} 0 \\ v+h(x^\mu ,y)\end{pmatrix} 
\end{equation}
with $\langle h \rangle = 0$ where $h$ is a real scalar field (Higgs field). We must of course have the nontrivial periodicity condition $\phi (x^\mu ,y+2\pi R) = B(2\pi R)\phi (x^\mu ,y)$ which implies
\begin{equation}
B^\dagger (2\pi R)U (x^\mu ,y + 2\pi R)B(2\pi R) = U (x^\mu ,y).
\end{equation} 
We then invoke the following gauge, 
\begin{equation}
W_A \rightarrow S W_A S^\dagger - \frac{i}{g}(\partial _A S)S^\dagger
\end{equation}
(where remember $W_\mu = \tilde{W}_\mu + \partial _\mu T$) where $S$ obeys the following differential equation,
\begin{equation}    
igS W_A S^\dagger + (\partial _A S)S^\dagger = B^\dagger [igUB W_A B^\dagger U^\dagger + (\partial _A U)U^\dagger ]B 
\end{equation}
where $S$ is unitary. 

Under the above gauge in (15) and the gauge condition in (16) we have 
\begin{equation}
D_A \phi = (\partial _A + igB W_A B^\dagger + \frac{i}{2}g'B_A) B \begin{pmatrix} \phi _1 \\ \phi _2 \end{pmatrix} \underset{\rm{gauge}}{\rightarrow} U D_A \phi , \nonumber
\end{equation}
and therefore 
\begin{equation}
{\rm{Tr}}(F_{AB}F^{AB}) \underset{\rm{gauge}}{\rightarrow}{\rm{Tr}}(F_{AB}F^{AB}). \nonumber
\end{equation}
There is no closed form analytical solution for $S$. Instead the general solution is a very cumbersome power series in embedded commutators of matrices. Thus the general solution of $S$ will not be derived or presented here. Notice that as $b$ goes to zero
\begin{equation}
S \underset{b \, = \, 0}{\rightarrow} U
\end{equation}  
which is the unitary gauge as expected. It should be noted that as $b$ goes to zero, equations (8) and (9) are no longer valid (they diverge in the zero mode limit).
\subsection{The masses and ratio relation}
With the above gauge we have for the masses of $\tilde{W} _\mu ^+$ and $\tilde{W} _\mu ^- $   
\begin{equation}
m_W = \sqrt{\frac{b^2}{R^2} + \frac{g^2 v^2}{4}} \, .
\end{equation}
The mass of the $Z$, the photon, and the remaining Higgs field are as expected since they are neutral. 
The phenomenologically forbidden 5th components for the gauge fields were delt with in section IIA. As previously pointed out in section IIA, these phenomenologically forbidden fifth components still carry their respective degrees of freedom in the model, they just don't interact with the other fields in the model and are thus phenomenologically hidden. The relation for the mass ratio is
\begin{equation}
\frac{m_W}{m_Z} = \sqrt{\frac{4 b^2}{v^2 R^2 (g^2 + g'^2)} + \frac{g^2 }{g^2 + g'^2}}.
\end{equation}
The appropriate standard model limits of the above expressions can be understood by noting that as the compactification size becomes very small, the flux is proportional to the area of the extra dimension as explained in section IIB.

We then find,  
\begin{equation}
m_{W_n} = \sqrt{\frac{(n+b)^2}{R^2} + \frac{g^2 v^2}{4}} \quad \rm{(n \; even)}\, , 
\end{equation}
for the mass of each mode of $\tilde{W} _\mu ^{+-} $. The modal masses for the $Z$ gauge boson, the photon, and the remaining Higgs field are as expected for this geometry.       
\subsection{Degrees of freedom in the model}
Notice that the kinetic energy term of $W_5 ^{+-}$ is hidden in the field combinations of equations (8) and (9) in the $(\partial _5 \tilde{W}_\mu ^- -\frac{ib}{R}\tilde{W}_\mu ^-)(\partial _5 \tilde{W}_\mu ^+ +\frac{ib}{R}\tilde{W}_\mu ^+)$ term of equation (10). Since the kinetic piece of the  $W_5 ^{+-}$ is hidden, so are the 2 degrees of freedom associated with the two fields $W_5 ^{+-}$. Therefore we have a total of 16 degrees of freedom before the Higgs mechanism in this higher dimensional space. After the Higgs mechanism we will have the term 
$\frac{1}{2}g^2 \begin{pmatrix} 0, & v\end{pmatrix}(\partial _\mu T)(\partial ^\mu T)\begin{pmatrix} 0 \\ v\end{pmatrix} = \frac{1}{4}g^2 v^2 \beta \Lambda (\partial _\mu W_5 ^+)(\partial ^\mu W_5 ^-)$ from (11) which is the kinetic term for the $W_5 ^{+-}$ modes after integrating out the extra coordinate (then $\beta \rightarrow \beta _n = -\frac{iR}{n+b}$ and $\Lambda \rightarrow \Lambda _n = \frac{iR}{n+b}$) thus restoring the 2 degrees of freedom associated with these two fields. The neutral weak gauge field $Z_\mu$ also acquires a mass after this mechanism and therefore gains 1 degree of freedom. So there is a total of 3 extra degrees of freedom that are balanced by the gauging away of the three Higgs fields in $U$. Thus we have 16 degrees of freedom before and after the Higgs mechanism. We now see why $\tilde{W}_\mu ^{+-}$ are the physical fields and not $W_\mu ^{+-}$.
\section{Addition of leptons and quarks to the model}
We add the following fermion terms to the Lagrangian density in (1): \\
$\EuScript{L}_{\rm{fermions}} = i\begin{pmatrix} \bar{\nu}^e _L , & \bar{e}_L \end{pmatrix}D_\mu \gamma ^\mu \begin{pmatrix} \nu ^e _L \\ e_L \end{pmatrix} + i\bar{e}_R D_\mu \gamma ^\mu e_R + i\begin{pmatrix} \bar{u}_L , & \bar{d}_L \end{pmatrix}D_\mu \gamma ^\mu \begin{pmatrix} u_L \\ d_L \end{pmatrix} \\ + i\bar{u}_R D_\mu \gamma ^\mu u_R +  i\bar{d}_R D_\mu \gamma ^\mu d_R - \begin{pmatrix} \bar{\nu}^e _L , & \bar{e}_L \end{pmatrix}D_5 \gamma ^5 \begin{pmatrix} \nu ^e _R \\ e_R \end{pmatrix} - \begin{pmatrix} \bar{u}_L , & \bar{d}_L \end{pmatrix}D_5 \gamma ^5 \begin{pmatrix} u_R \\ d_R \end{pmatrix} \\ - \lambda _e \begin{pmatrix} \bar{\nu}^e _L , & \bar{e}_L \end{pmatrix}\phi e_R - \lambda _d \begin{pmatrix} \bar{u}_L , & \bar{d}_L \end{pmatrix}\phi d_R - \lambda _u \begin{pmatrix} \bar{u}_L , & \bar{d}_L \end{pmatrix}i\tau ^2 \phi ^* u_R$ where $i\tau ^2 = \begin{pmatrix} 0 & 1 \\ -1 & 0 \end{pmatrix}$. It is understood that in this notation, the right handed singlets do not couple to the SU(2) gauge fields in the covariant derivatives. The two $\gamma ^5$ terms that couple right handed doublets to left handed doublets are necessary because they give the modal and flux dependence of the masses for the fermion modes. These two terms are non chiral, however in the standard model limit (zero mode and zero flux limit) these two terms vanish in our orbifold geometry thus restoring the chirality of the theory. Notice that we now have nontrivial couplings of the Higgs and lepton fields to the gauge bosons due to the fluxes. Please see [14] for the details of these new couplings.

For the fermion spinors we have,
\begin{equation}
\begin{pmatrix} \nu ^e _L \\ e_L \end{pmatrix} \underset{\rm{flux}}{\rightarrow} \begin{pmatrix} 1 & 0 \\ 0 & e^{-iby/R}\end{pmatrix} \begin{pmatrix} \nu ^e _L \\ e_L \end{pmatrix} \, ,
\end{equation}
\begin{equation}
\begin{pmatrix} u _L \\ d_L \end{pmatrix} \underset{\rm{flux}}{\rightarrow} \begin{pmatrix} e^{\frac{2}{3}iby/R} & 0 \\ 0 & e^{\frac{1}{3}iby/R}\end{pmatrix} \begin{pmatrix} u _L \\ d_L \end{pmatrix} \, ,
\end{equation}
and the same for the right handed fermion spinors where $u_R \underset{\rm{flux}}{\rightarrow} e^{\frac{2}{3}iby/R} u_R$, 
$d_R \underset{\rm{flux}}{\rightarrow} e^{\frac{1}{3}iby/R} d_R$ and $e_R \underset{\rm{flux}}{\rightarrow} e^{-iby/R} e_R$. From (13) we had
\begin{equation} 
\phi (x^\mu ,y) = \frac{1}{\sqrt{2}}U(x^\mu ,y)B(y)\begin{pmatrix} 0 \\ v+h(x^\mu ,y)\end{pmatrix}. \nonumber
\end{equation}
In conjunction with (15), with $S$ obeying (16), we have for the fermion spinors,
\begin{equation}
\begin{pmatrix} 1 & 0 \\ 0 & e^{-iby/R}\end{pmatrix}\begin{pmatrix} \nu^e _L  \\ e_L \end{pmatrix}\rightarrow U \begin{pmatrix} 1 & 0 \\ 0 & e^{-iby/R}\end{pmatrix}\begin{pmatrix} \nu^e _L  \\ e_L \end{pmatrix} \, , 
\end{equation}
\begin{equation}
\begin{pmatrix} e^{\frac{2}{3}iby/R} & 0 \\ 0 & e^{\frac{1}{3}iby/R}\end{pmatrix}\begin{pmatrix} u _L \\ d_L \end{pmatrix}\rightarrow U \begin{pmatrix} e^{\frac{2}{3}iby/R} & 0 \\ 0 & e^{\frac{1}{3}iby/R}\end{pmatrix}\begin{pmatrix} u _L \\ d_L \end{pmatrix} \, ,  
\end{equation}
and the same for the right handed spinors. Thus, after some matrix algebra, we have as expected, $m_e = \lambda _e \sqrt{\frac{v^2}{2}}$, $m_u = \lambda _u \sqrt{\frac{v^2}{2}}$ and $m_d = \lambda _d \sqrt{\frac{v^2}{2}}$ for the fermion masses in the bulk.

For the quark masses in the brane, we have the following terms after integrating out the extra coordinate for the $n$th even mode,
\begin{equation}  
-\begin{pmatrix} \bar{u} _{L_n} ,  & \bar{d}_{L_n} \end{pmatrix}[\begin{pmatrix} in & 0 \\ 0 & in\end{pmatrix} + \begin{pmatrix} \frac{2}{3}\frac{ib}{R} & 0 \\ 0 & \frac{1}{3}\frac{ib}{R}\end{pmatrix}]\gamma ^5 \begin{pmatrix} u _{R_n} \\ d_{R_n} \end{pmatrix} - \lambda _u \sqrt{\frac{v^2}{2}} \bar{u} _{L_n} u _{R_n} - \lambda _d \sqrt{\frac{v^2}{2}} \bar{d}_{L_n} d_{R_n} \, , \nonumber 
\end{equation}
where the first term comes from 
\begin{equation}
- \begin{pmatrix} \bar{u}_L , & \bar{d}_L \end{pmatrix}\begin{pmatrix} e^{-\frac{2}{3}iby/R} & 0 \\ 0 & e^{-\frac{1}{3}iby/R}\end{pmatrix}D_5 \gamma ^5 \begin{pmatrix} e^{\frac{2}{3}iby/R} & 0 \\ 0 & e^{\frac{1}{3}iby/R}\end{pmatrix}\begin{pmatrix} u_R \\ d_R \end{pmatrix}. \nonumber 
\end{equation} 
If we then let, 
\begin{equation}
u_n \rightarrow e^{i\alpha _n \gamma ^5}u_n \quad {\rm{where}} \quad e^{2i\alpha _n \gamma ^5} = \cos{2\alpha _n} + i\gamma ^5 \sin{2\alpha _n} = \frac{\lambda _u v}{m_{u_n}} - i\gamma ^5 \frac{n+\frac{2}{3}b}{m_{u_n}R} 
\end{equation}
and
\begin{equation}
d_n \rightarrow e^{i\sigma _n \gamma ^5}d_n \quad {\rm {where}} \quad e^{2i\sigma _n \gamma ^5} = \cos{2\sigma _n} + i\gamma ^5 \sin{2\sigma _n} = \frac{\lambda _d v}{m_{d_n}} - i\gamma ^5 \frac{n+\frac{1}{3}b}{m_{d_n}R} \, ,  
\end{equation}
we find
\begin{equation}
m_{u_n} = \sqrt{\frac{\lambda _u ^2 v^2}{2} + \frac{(n+\frac{2}{3}b)^2}{R^2}} \quad \rm{(n \; even)}\, ,
\end{equation}
\begin{equation}
m_{d_n} = \sqrt{\frac{\lambda _d ^2 v^2}{2} + \frac{(n+\frac{1}{3}b)^2}{R^2}} \quad \rm{(n \; even)}.
\end{equation}

Similarly if we let,
\begin{equation} 
e_n \rightarrow e^{i\beta _n \gamma ^5}e_n \quad {\rm {where}} \quad e^{2i\beta _n \gamma ^5} = \cos{2\beta _n} + i\gamma ^5 \sin{2\beta _n} = \frac{\lambda _e v}{m_{e_n}} - i\gamma ^5 \frac{n-b}{m_{e_n}R}
\end{equation}
and 
\begin{equation}
\nu_{e_n} \rightarrow i\gamma ^5 \nu_{e_n} \, ,
\end{equation}
we find 
\begin{equation}
m_{e_n} = \sqrt{\frac{\lambda _e ^2 v^2}{2} + \frac{(n-b)^2}{R^2}} \quad \rm{(n \; even)} \, ,
\end{equation}
\begin{equation}
m_{{\nu_e}_n} = \frac{|n|}{R} \quad \rm{(n \; even)}.
\end{equation}
The transformations in (25),(26),(29), and (30) are required to get physical mass terms for the fermion modes as well as re-establish the Ward-Takahashi identity [11]. For further mathematical detail, please see [14].

In QED, the interaction term $\sum _{n,m \, = \, -\infty}^\infty \bar{\psi}_n \gamma ^\mu \psi _m A_{\mu , n-m}$, as well as the coupling terms for the fermions and the weak gauge bosons along the spacetime directions, are actually non invariant under the above transformations (except for $n=m$) for the fermion modes. Thus, as a result of the flux, QED becomes parity violating in this theory [11]. 
\section{Rough Estimate of Upper limit on $|\frac{b}{R}|$}   
From equation (19) we have  
\begin{equation}
\frac{m_W ^2}{m_Z ^2} = \cos ^2 {\theta _W} + \frac{4 b^2 \cos ^2 {\theta _W}}{g^2 v^2 R^2} \nonumber
\end{equation}
where $\cos {\theta _W} = \frac{g^2}{g^2 + g'^2}$ at tree level in the standard model.
Then 
\begin{equation}
\frac{m_W ^2}{m_Z ^2 cos ^2 {\theta _W}} = 1 + \frac{4 b^2}{g^2 v^2 R^2} \nonumber
\end{equation}
and since $m_W = \frac{gv}{2}$ in the standard model (not to be confused with $m_W$ from our model) we obtain
\begin{equation}
\frac{m_W ^2}{m_Z ^2 cos ^2 {\theta _W}} = 1 + \frac{ b^2}{m_W ^2 R^2}. \nonumber
\end{equation}
Theoretically at tree level $\frac{m_W}{m_Z cos{\theta _W}} = 1$ in the standard model, but if radiative corrections are included, then this is changed to $\frac{m_W ^2}{m_Z ^2 cos ^2 {\theta _W}}(1 + \frac{\alpha T}{2})^2$ [5]. Finally,
\begin{equation}
\frac{m_W ^2}{m_Z ^2 cos ^2 {\theta _W}}\Bigg (1 + \frac{\alpha T}{2}\Bigg ) ^2 = 1 + \frac{ b^2}{m_W ^2 R^2}
\end{equation}
and thus we can associate $\frac{ b^2}{m_W ^2 R^2}$ as the deviation to the standard model due to the extra dimensional flux. The quantities $\frac{m_W ^2}{m_Z ^2 cos ^2 {\theta _W}}$ and $(1 + \frac{\alpha T}{2})^2$ have experimental values and error associated with their values, namely $\frac{m_W ^2}{m_Z ^2 cos ^2 {\theta _W}}=1.0111 \pm 0.00089$ and $(1 + \frac{\alpha T}{2})^2=1.0002 \pm 0.00070$ [5]. So we can find the experimental value and experimental error in the quantity $\frac{m_W ^2}{m_Z ^2 cos ^2 {\theta _W}}(1 + \frac{\alpha T}{2})^2$ using basic error analysis. We find
\begin{equation}
\frac{m_W ^2}{m_Z ^2 cos ^2 {\theta _W}}\Bigg (1 + \frac{\alpha T}{2}\Bigg )^2 = 1.0113 \pm 0.0016 \,[5].
\end{equation}
We can then attribute the upper limit for $|\frac{b}{R}|$ to the experimental error in the quantity $\frac{m_W ^2}{m_Z ^2 cos ^2 {\theta _W}}(1 + \frac{\alpha T}{2})^2$ or
\begin{equation}
\frac{ b^2}{m_W ^2 R^2} < 0.0016 
\end{equation}
which finally gives 
\begin{equation}
\Bigg |\frac{b}{R}\Bigg | < 3.41 \, \rm {GeV}.   
\end{equation}
Using $(1 + \frac{\alpha T}{2})^2=0.9994 \pm 0.00090$ [5] gives
\begin{equation}
\Bigg |\frac{b}{R}\Bigg | < 3.22 \, \rm {GeV}.   
\end{equation}

We have applied the theory to observable phenomena to get upper limits for well known quantities such as the magnetic dipole moment and the electric dipole moment of charged leptons [6,7,8,9,10] to first order in radiative corrections with the fluxes present in [11]. The results in [11] are in complete agreement with equations (36) and (37). As mentioned before, these fluxes could come from superconducting cosmic strings [12]. Indeed, a current from a light charged lepton moving at a distance $\sim R$ from one of these strings of about one ampere (section V of [11]) could produce the flux associated with the upper limit of the ratio of $b$ and $R$ in our model with the current accepted upper limit on $R$ [13] for this type of geometry.

\end{document}